# Superconducting fiber with Tc up to 7.43 K in $Nb_2Pd_xS_{5-\delta}$ (0.6<x<1)

Hongyan Yu,[†] Ming Zuo,[‡] Lei Zhang,[†] Shun Tan,[‡] Changjin Zhang,[*,†,‡] and Yuheng Zhang[†,‡]

[†]High Magnetic Field Laboratory, Chinese Academy of Sciences and University of Science and Technology of China, Hefei 230026, People's Republic of China

[‡]Hefei National Laboratory for Physical Sciences at Microscale, University of Science and Technology of China, Hefei 230026, People's Republic of China

*Supporting Information Placeholder*

**ABSTRACT:** Wiring systems powered by high-efficient superconductors have long been a dream of scientists, but researchers have faced practical challenges such as finding flexible materials. Here we report superconductivity in $Nb_2Pd_xS_{5-\delta}$ fibers with transition temperature up to 7.43 K, which have typical diameters of 0.3~3 μm. Superconductivity occurs in a wide range of Pd (0.6<x<1) and S (0<δ<0.61) contents, suggesting that the superconductivity in this system is very robust. Long fibers with suitable size provide a new route to high-power transmission cables and electronic devices.

Exploring new superconducting materials has long been a topic of interest in condensed matter physics and material science communities as well as industrial field. Cuprate superconductors[1] and iron-based superconductors[2] are two kinds of particularly important superconducting materials among the hundreds of superconducting materials discovered up to now. The highest superconducting transition temperature ($T_c$) at ambient pressure can be 132 K and 56 K for cuprate superconductors and iron-based superconductors, respectively. Some high temperature superconductors have become practically applicable in recent years. For example, high-quality superconducting wires based on $YBa_2Cu3O_{7-\delta}$ and $Bi_2Sr_2CaCu_2O_8$ superconductors have been fabricated for demonstrations of superconducting magnets, transmission cables, motors and other electrical power components.[3] However, these superconducting materials are both ceramic, so that they are difficult to be shaped. For practical uses, it is very important that a superconducting material is flexible. In recent years, new superconducting compounds, such as $BaTi_2Sb_2O$, $Bi_4O_4S_3$, $Nb_2Pd_{0.81}S_5$, $Nb_3Pd_{0.7}S_7$, $Li_2Pd_3B$, $LaNiGa_2$, and $Ir_xCh_2$ (Ch=Se and Te) are discovered.[4-11] Despite of the relatively lower superconducting transition temperature (less than 10 K), their discovery attracts much attention due to their potential application perspectives and scientific importance. Here we report the growth and the characterization of $Nb_2Pd_xS_{5-\delta}$ (0.6<x<1) superconducting fibers by using traditional flux-melting method. The fibers have typical diameters of 0.3~3 μm. The Tc value can be 7.43 K in $Nb_2Pd_{0.963}S_{4.967}$. It is found that within a wide range of Pd-site and S-site occupancy rates (0.6<x< 1, δ<0.61) the $Nb_2Pd_xS_{5-\delta}$ fibers exhibit superconducting transition, suggesting the superconductivity in $Nb_2Pd_xS_{5-\delta}$ fibers is very robust.

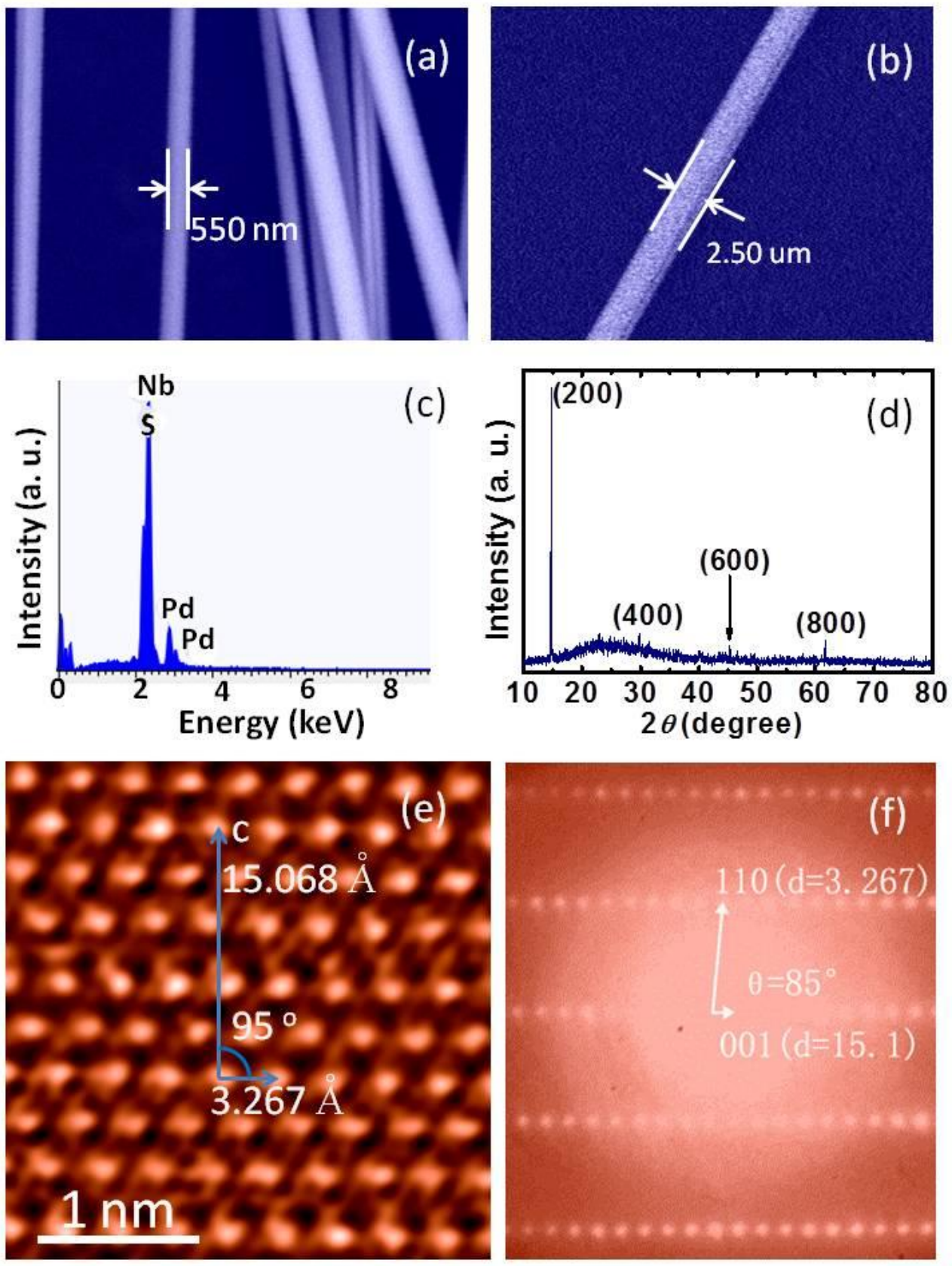

***Figure 1.*** Crystal structure and chemical content. a) and b) SEM images of the as-grown fibers and a single fiber, respectively. c) The EDX spectrum of a single fiber. d) XRD pattern of the as-grown fibers. e) Atomic-resolution TEM image of the fiber taken along the [1$\bar{1}$0] zone-axis direction. f) HRTEM electron diffraction pattern of e).

Figure 1a shows a typical scanning electron microscopy (SEM) image of the as-grown fibers. They are cylindrical in shape and the length can be as long as 4 cm (Figures S1 and S2). Figure 1b gives the SEM image of a single fiber. The diameter of the fiber shown in Figure 1b is about 2.5 μm. It is found that these fibers can easily be bent into any shape, suggesting that they are flexible. In order to determine the chemical contents of the fibers, we perform an energy dispersive x-ray spectroscopy analysis (EDX) on these samples. The EDX spectrum (Figure 1c) confirms the presence of Nb, Pd, and S. A comparison between the starting compositions and the actual compositions of the fibers grown under different conditions is listed in Table S1. As can be seen from Table S1, excess S is added in the starting materials in order to compensate the evaporation at high temperature. It should be noticed that the occupancy rate of Pd is generally less than 1, meaning that there is some Pd-site vacancy in the fibers. From the chemical analysis we also notice that there is a certain vacancy at S-site.

X-ray diffraction patterns of the as-grown $Nb_2Pd_xS_{5-\delta}$ ($0.6<x<1$) fibers (Figure 1d) taken on the surface of the cylinder confirm the single crystal behavior of the fiber. We notice that the chemical formula of $Nb_2Pd_xS_{5-\delta}$ is very similar to that of previous-discovered $Nb_2Pd_{0.71}S_5$ and $Nb_3Pd_{0.72}Se_7$.[12,13] Using the same monoclinic *C2/m* space group and similar lattice parameters we find that the diffraction peaks shown in Figure 1d can be indexed to the diffraction from the ($l$00) planes. And the *a*-axis lattice constant is determined to be 12.041 Å. From the atomic-resolution tunneling electron microscopy (TEM) image and the high resolution tunneling electron microscopy (HRTEM) electron diffraction pattern (Figure 1e and 1f, Figure S3) results the *b*- and *c*-axis lattice constants are determined to be 3.396 Å and 15.068 Å, respectively.

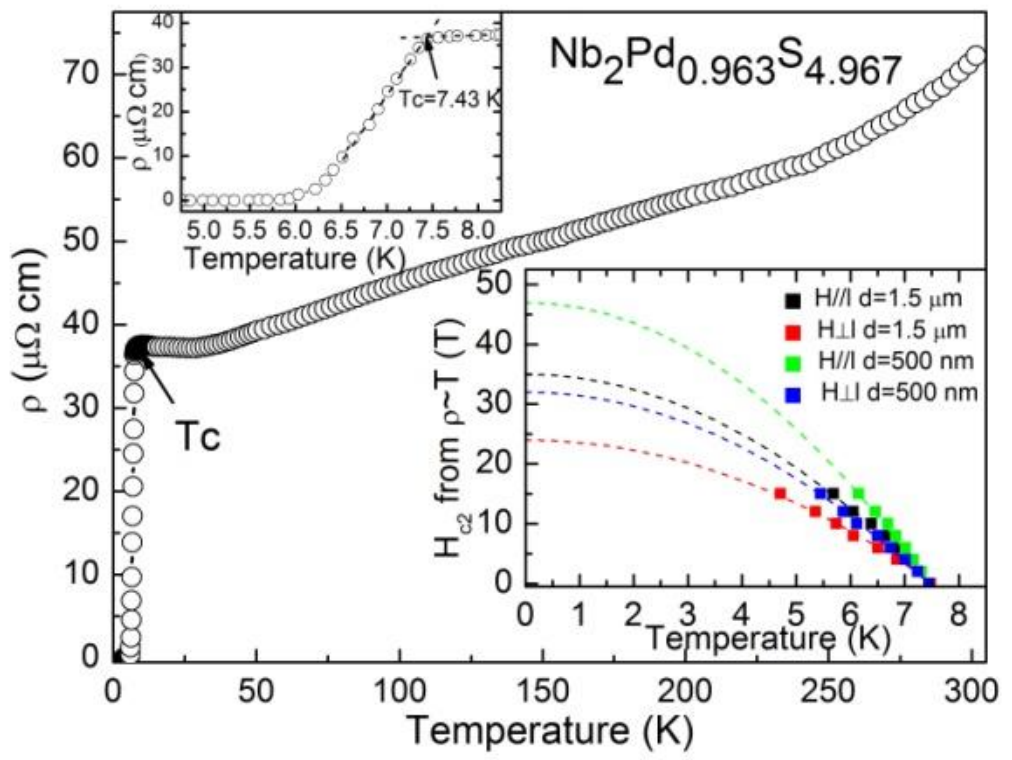

***Figure 2.*** Temperature dependence of resistivity of the $Nb_2Pd_{0.963}S_{4.967}$ fiber. The upper-left inset shows an enlarged view near the transition temperature. The inset at lower right corner gives the upper critical field determined from the magneto-transport measurements. The dashed curves are the fitting data using $H_{c2}(T)= H_{c2}(0)\{1-(T/Tc)^2\}$.

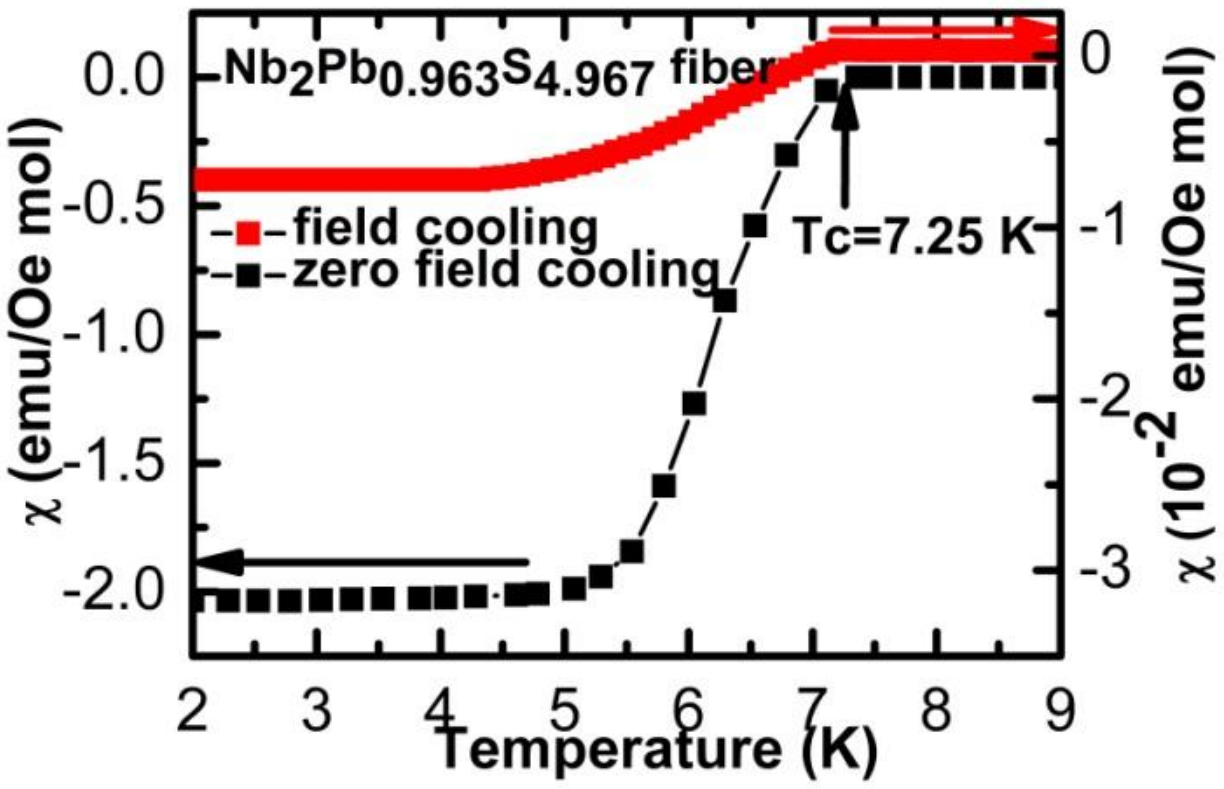

***Figure 3.*** Temperature dependence of magnetic susceptibility of the $Nb_2Pd_{0.963}S_{4.967}$ fiber measured under both zero field cooling process and field cooling process. The inset shows the magnetic susceptibility at high temperature region.

Figure 2 gives the temperature dependence of electrical resistivity (ρ~T curve) of the $Nb_2Pd_{0.963}S_{4.967}$ fiber. The ρ~T curve is measured on one single fiber by using four-probe method. At high temperature, the ρ~T curve exhibits metallic-like behavior. Clear resistivity drop can be observed below Tc~7.43 K. In order to verify that this resistivity drop is originated from superconducting transition, we perform measurements of magnetic susceptibility on the fibers. Figure 3 gives the temperature dependence of magnetic susceptibility of the $Nb_2Pd_{0.963}S_{4.967}$ superconducting fiber. At high temperature region, the magnetic susceptibility slightly increases with decreasing temperature without showing any signature of magnetic phase transition. Below ~7.2 K, diamagnetic signal presents, indicating the occurrence of superconducting transition. It can be seen that the superconducting transition width is less than 2 K, suggesting the high quality of the fiber. The ρ~T curves measured under magnetic field suggest that the transport upper critical field is strongly dependent on the diameter of the fiber (Figure S4, the lower right inset of Figure 2). The fibers with smaller diameter exhibit higher upper critical field. Similarly, the critical current density ($J_c$) exhibits the same dependence on the diameter of the fiber (Figure S5). These facts suggest that both the transport upper critical field and $J_c$ can be substantially enhanced by decreasing the diameter of the fiber.

We compare the Tc values for the samples prepared using different growth conditions and different starting compositions (Table S1). We find that both slow-cooling process and quenching method can yield superconducting fibers. The fibers prepared by quenching method exhibit higher transition temperature compared with those prepared by slow-cooling process. Furthermore, it is found that the fibers grown at 800° C (the maximum temperature) followed by quenching down to room temperature exhibit higher Tc value. The Pd-site occupancy rate can be adjusted by changing the initial Pd content in the starting materials. Generally, we can obtain higher Pd-site occupancy rate if we add more Pd in the nominal composition. Table S1 suggests that the Pd-site occupancy rate varies from 0.6 to 1. It is interesting to notice that all $Nb_2Pd_xS_{5-\delta}$ ($0.6<x<1$, $0<\delta<0.62$) fibers exhibit superconducting transition, indicating that the superconductivity in this system is quite robust. The vacancy at Pd-site and S-site is reminiscent of the K- and Fe-site vacancy in the $K_xFe_{2-y}Se_2$ superconductor.[14] However, the role of the Pd- and S-site vacancy in the occurrence of superconductivity in the $Nb_2Pd_xS_{5-\delta}$ fibers needs further investigation.

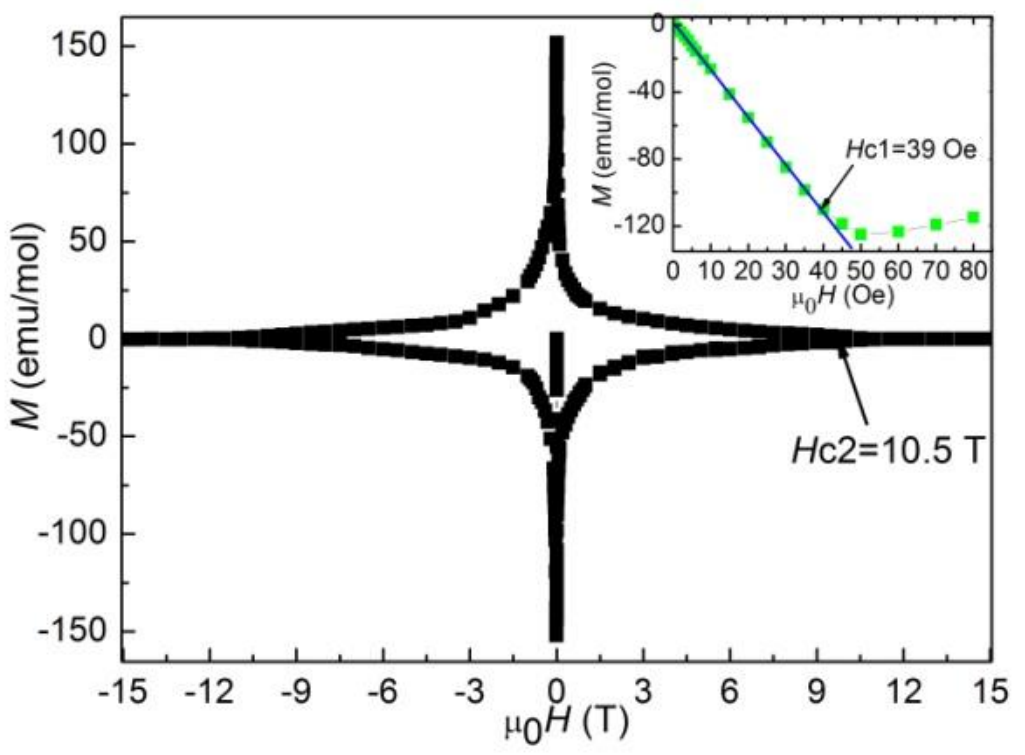

***Figure 4.*** Magnetic field dependence of magnetic susceptibility of the $Nb_2Pd_{0.963}S_{4.967}$ fiber measured at 2 K. The green curve gives the initial $M \sim \mu_0 H$ isotherm. The blue line gives a linear fitting at low field region.

Figure 4 shows the magnetic hysteresis loop of the $Nb_2Pd_{0.963}S_{4.967}$ fiber measured at 2 K. The $M \sim \mu_0 H$ curve of a typical type-II superconductor is observed. The lower critical field ($H_{c1}$), defined as the field at which the $M \sim \mu_0 H$ curve deviates from the initial linear slope, is 39 Oe at 2 K. The upper critical field ($H_{c2}$) at 2 K is about 10.5 T (105000 Oe). According to Ginzburg-Landau (GL) theory, the lower and upper critical fields at 0 K can be deduced using $H_c(T)= H_c(0)(1-(T/T_c)^2)$. Thus the $H_{c1}(0)$ and $H_{c2}(0)$ values are 42.3 Oe and 113200 Oe, respectively. The GL parameter $\kappa(0)$ is estimated to be 76 using the equation $H_{c2}(0)/H_{c1}(0)=2\kappa(0)^2/\ln\kappa(0)$.[15]

The GL coherence length at 0 K ($\xi(0)$), estimated using $\xi(0)=\{\Phi_0/(2\pi H_{c2}(0))\}^{1/2}$, is about 5.4 nm. And the GL penetration depth $\lambda(0)$ is 410 nm based on the equation $\kappa(0)=\lambda(0)/\xi(0)$. It should be noted that the $\lambda(0)$ value is comparable with the diameters of the $Nb_2Pd_xS_{5-\delta}$ superconducting fibers, resulting in the fact that magnetic field can nearly completely penetrate through the fiber samples. In this sense, the $Nb_2Pd_xS_{5-\delta}$ fibers can serve as a possible candidate material to accurately measure the penetration depth.

In conclusion, we have found that superconductivity with Tc up to 7.43 K occurs in $Nb_2Pd_xS_{5-\delta}$ single crystal fibers. Within a wide range of Pd (0.6<x<1) and S (0<δ<0.62) contents the fibers exhibit superconducting transition, suggesting that the superconductivity in this system is rather robust. The present results demonstrate that it is possible to fabricate flexible superconducting fibers for industrial applications.

## ASSOCIATED CONTENT

### Supporting Information

Experimental details, synthesis conditions of $Nb_2Pd_xS_{5-\delta}$ (0.6<x<1) fibers, supporting table and figures are included in the supporting information. This material is available free of charge via the Internet at http://pubs.acs.org.

## AUTHOR INFORMATION

### Corresponding Author

zcjin@ustc.edu.cn

Notes

The authors declare no competing financial interest.

## ACKNOWLEDGMENT

This work was supported by the State Key Project of Fundamental Research of China (Grant Nos. 2010CB923403 and 2011CBA00111), the Nature Science Foundation of China (Grant Nos. 11174290 and U1232142), and the Hundred Talents Program of the Chinese Academy of Sciences.

# Supporting information

# Superconducting fiber with Tc up to 7.43 K in $Nb_2Pd_xS_{5-\delta}$ (0.6<x<1)

Hongyan Yu,[†] Ming Zuo,[‡] Lei Zhang,[†] Shun Tan,[‡] Changjin Zhang,[*,†,‡] and Yuheng Zhang[†,‡]

[†]High Magnetic Field Laboratory, Chinese Academy of Sciences and University of Science and Technology of China, Hefei 230026, People's Republic of China

[‡]Hefei National Laboratory for Physical Sciences at Microscale, University of Science and Technology of China, Hefei 230026, People's Republic of China

*Index* *Page*

## Sample preparations and experimental details

**Sample preparations**. The $Nb_2Pd_xS_{5-\delta}$ (0.6<x<1) fibers were prepared from a reaction of the elements combined in the ratios Nb:Pd:S listed in Table S1 (Nb powder, 99.9%, ALFA AESAR; Pd powder, 99.9%, ALFA AESAR; S powder, 99.5%, ALFA AESAR). The metals Nb and Pd were reduced under a flow of $H_2$ at 600° C for 5 hours prior to reaction. The elements were loaded into a quartz tube in an Argon-filled glove box (BRAUN MBC-Unilab). After taking out from the glove box, the quartz tube was evacuated to $10^{-6}$ torr and sealed. The mixture was placed in a high temperature furnace that was heated from 30°C to the maximum temperature (750-850°C) at a rate of 10°C/h. After being heated to the maximum temperature for 12 hours, some of the quartz tubes were removed from the furnace and quenched down to 0°C by putting the quartz tubes into water ice. In order to compare the resultant transition temperature, we also prepared some samples using different cooling down process. These samples were cooled down from the maximum temperature to 500°C at a rate of 3°C/h, which was followed by cooling down to room temperature at a rate of 50°C/h.

**Scanning electron microscopy and Energy dispersive x-ray spectroscopy.** Scanning electron microscopy measurement was performed using a Hitachi TM3000 STM. Typical amplification factors used in the SEM images of the $Nb_2Pd_xS_{5-\delta}$ fibers are 5000, 10000, and 20000. All samples are cylindrical in shape. The diameters of the $Nb_2Pd_xS_{5-\delta}$ fibers can be accurately measured. The smallest fiber we obtained has diameter of 0.285 μm and the largest one has diameter of 3 μm. The energy dispersive x-ray spectroscopy (EDX) analysis was performed using Oxford SWIFT3000 spectroscopy equipped with a Si detector. The EDX measurements were done on single fibers. For each fiber samples, at least ten samples were measured. The error bars are indicated in Table S1.

**High resolution tunneling electron microscopy and atomic resolution tunneling electron microscopy.** High resolution tunneling electron microscopy (HRTEM) measurement was performed using a JEOL-2010 transmission electron microscope. The point-to-point resolution is better than 0.2 nm. Prior to the HRTEM measurement, the $Nb_2Pd_xS_{5-\delta}$ fiber samples were ground into fine powder specimens. The specimens were then loaded into a copper grid which serves as the sample holder during the HRTEM measurement. The applied accelerating voltage is 200 kV in the measurement. Both the HRTEM images of the specimens and the electron-diffraction patterns were taken. Atomic resolution tunneling electron microscopy measurement was performed using a JEOL-ARM-200F microscope which offers resolution of 0.08 nm at 200 kV. The operational procedure is similar to the HRTEM measurement.

**Table S1. The nominal and real compositions of the $Nb_2Pd_xS_{5-\delta}$ fibers grown under different maximum reaction temperatures and different ways in the cooling down process and the corresponding superconducting transition temperature ($T_c$)**

| Nominal composition | Real composition | Maximum temperature | Cooling down method | Tc (K) |
|---|---|---|---|---|
| $Nb_2PdS_{4.5}$ | $Nb_2Pd_{0.887(12)}S_{4.388(20)}$ | 850° C | quench* | 4.67 |
| $Nb_2PdS_5$ | $Nb_2Pd_{0.902(12)}S_{4.684(19)}$ | 850° C | quench | 4.95 |
| $Nb_2PdS_6$ | $Nb_2Pd_{0.919(11)}S_{4.855(11)}$ | 850° C | quench | 6.28 |
| $Nb_2PdS_{6.5}$ | $Nb_2Pd_{0.917(09)}S_{4.880(12)}$ | 850° C | quench | 6.05 |
| $Nb_2PdS_7$ | $Nb_2Pd_{0.908(08)}S_{4.849(14)}$ | 850° C | quench | 5.93 |
| $Nb_2PdS_8$ | $Nb_2Pd_{0.905(13)}S_{4.841(14)}$ | 850° C | quench | 6.06 |
| $Nb_2Pd_{0.6}S_6$ | $Nb_2Pd_{0.604(05)}S_{4.781(10)}$ | 850° C | quench | 5.04 |
| $Nb_2Pd_{0.7}S_6$ | $Nb_2Pd_{0.692(08)}S_{4.811(09)}$ | 850° C | quench | 5.28 |
| $Nb_2Pd_{0.8}S_6$ | $Nb_2Pd_{0.792(08)}S_{4.845(10)}$ | 850° C | quench | 6.1 |
| $Nb_2Pd_{0.9}S_6$ | $Nb_2Pd_{0.829(11)}S_{4.841(16)}$ | 850° C | quench | 6.28 |
| $Nb_2Pd_{1.1}S_6$ | $Nb_2Pd_{0.948(08)}S_{4.843(13}$ | 850° C | quench | 7.08 |
| $Nb_2Pd_{1.2}S_6$ | $Nb_2Pd_{0.950(12)}S_{4.830(12)}$ | 850° C | quench | 6.93 |
| $Nb_2Pd_{1.3}S_6$ | $Nb_2Pd_{0.958(19)}S_{4.847(12)}$ | 850° C | quench | 6.91 |
| $Nb_2Pd_{1.4}S_6$ | $Nb_2Pd_{0.984(13)}S_{4.805(11)}$ | 850° C | quench | 6.58 |
| $Nb_2Pd_{1.5}S_6$ | $Nb_2Pd_{0.992(11)}S_{4.774(16)}$ | 850° C | quench | 6.53 |
| $Nb_2Pd_{0.6}S_6$ | $Nb_2Pd_{0.596(05)}S_{4.884(11)}$ | 800° C | quench | 5.05 |
| $Nb_2Pd_{0.7}S_6$ | $Nb_2Pd_{0.689(05)}S_{4.944(06)}$ | 800° C | quench | 5.26 |
| $Nb_2Pd_{0.8}S_6$ | $Nb_2Pd_{0.781(07)}S_{4.956(08)}$ | 800° C | quench | 5.31 |
| $Nb_2Pd_{0.9}S_6$ | $Nb_2Pd_{0.903(05}S_{4.943(06)}$ | 800° C | quench | 6.53 |
| $Nb_2PdS_6$ | $Nb_2Pd_{0.944(09)}S_{4.972(08)}$ | 800° C | quench | 7.06 |
| $Nb_2Pd_{1.1}S_6$ | $Nb_2Pd_{0.963(07)}S_{4.967(06)}$ | 800° C | quench | 7.43 |
| $Nb_2Pd_{1.2}S_6$ | $Nb_2Pd_{0.977(07)}S_{4.965(06)}$ | 800° C | quench | 7.11 |
| $Nb_2Pd_{1.3}S_6$ | $Nb_2Pd_{0.991(09)}S_{4.988(08)}$ | 800° C | quench | 6.45 |
| $Nb_2Pd_{1.4}S_6$ | $Nb_2Pd_{0.988(11)}S_{4.979(08)}$ | 800° C | quench | 5.86 |
| $Nb_2Pd_{1.5}S_6$ | $Nb_2Pd_{0.989(11)}S_{4.995(09)}$ | 800° C | quench | 5.02 |
| $Nb_2PdS_{4.5}$ | $Nb_2Pd_{0.909(10)}S_{4.414(17)}$ | 800° C | quench | 5.13 |
| $Nb_2PdS_5$ | $Nb_2Pd_{0.912(11)}S_{4.733(12)}$ | 800° C | quench | 5.65 |
| $Nb_2PdS_{6.5}$ | $Nb_2Pd_{0.939(06)}S_{4.941(10)}$ | 800° C | quench | 6.45 |
| $Nb_2PdS_7$ | $Nb_2Pd_{0.916(14)}S_{4.977(10)}$ | 800° C | quench | 6.27 |
| $Nb_2PdS_8$ | $Nb_2Pd_{0.911(11)}S_{4.989(12)}$ | 800° C | quench | 5.81 |
| $Nb_2PdS_6$ | $Nb_2Pd_{0.951(08)}S_{4.959(03)}$ | 800° C | slow cooling down※ | 6.26 |
| $Nb_2Pd_{1.1}S_6$ | $Nb_2Pd_{0.969(08)}S_{4.951(04)}$ | 800° C | slow cooling down | 6.49 |
| $Nb_2Pd_{1.2}S_6$ | $Nb_2Pd_{0.972(06)}S_{4.933(04)}$ | 800° C | slow cooling down | 5.66 |
| $Nb_2PdS_6$ | $Nb_2Pd_{0.933(08)}S_{4.943(06)}$ | 750° C | quench | 5.63 |
| $Nb_2Pd_{1.1}S_6$ | $Nb_2Pd_{0.937(08)}S_{4.937(08)}$ | 750° C | quench | 5.44 |
| $Nb_2Pd_{1.2}S_6$ | $Nb_2Pd_{0.941(13)}S_{4.943(08)}$ | 750° C | quench | 5.07 |

*Quench here means that the samples are taken out from the high temperature furnace at the maximum temperature (850°C, 800°C, or 750°C) and put into water ice shortly. ※slow cooling down means the samples are cooled down from 800°C to 500°C at a rate of 3 K/hour then followed by cooling down to room temperature at a rate of 50 K/hour.

Supplementary Figures

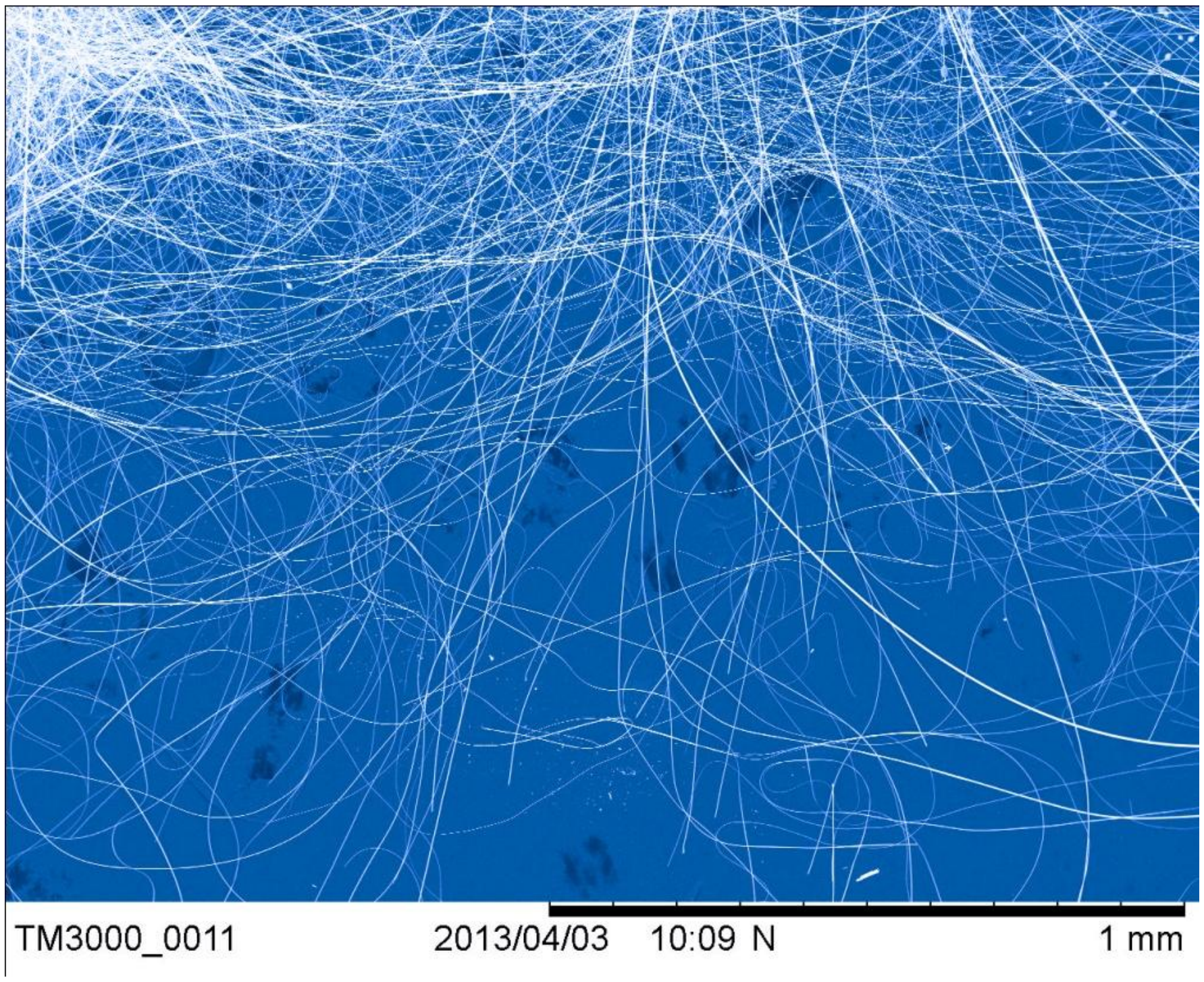

Figure S1. Scanning electron microscopy image shown the morphology of the as-grown fibers.

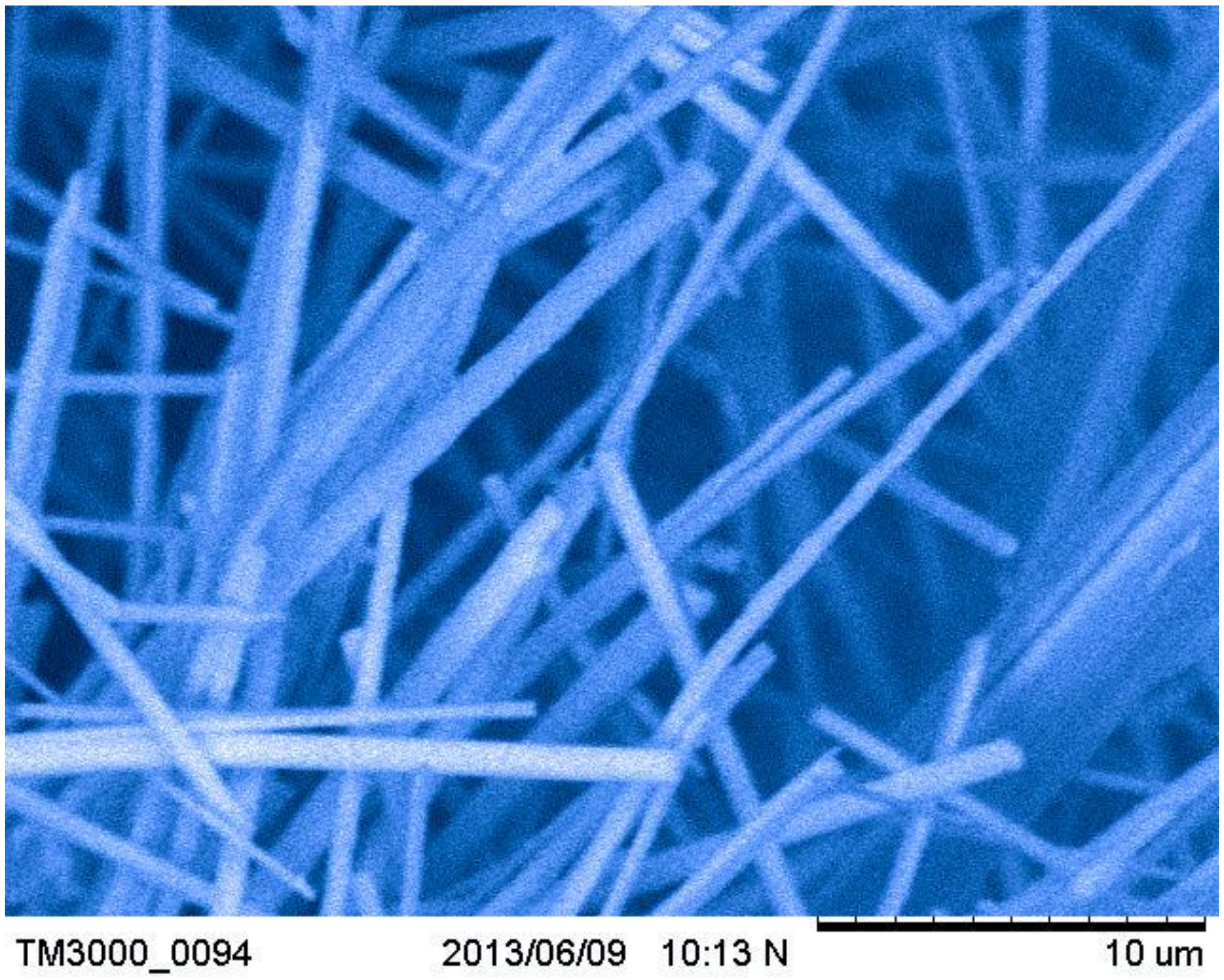

Figure S2. Scanning electron microscopy image shown the cross-section of the as-grown fibers

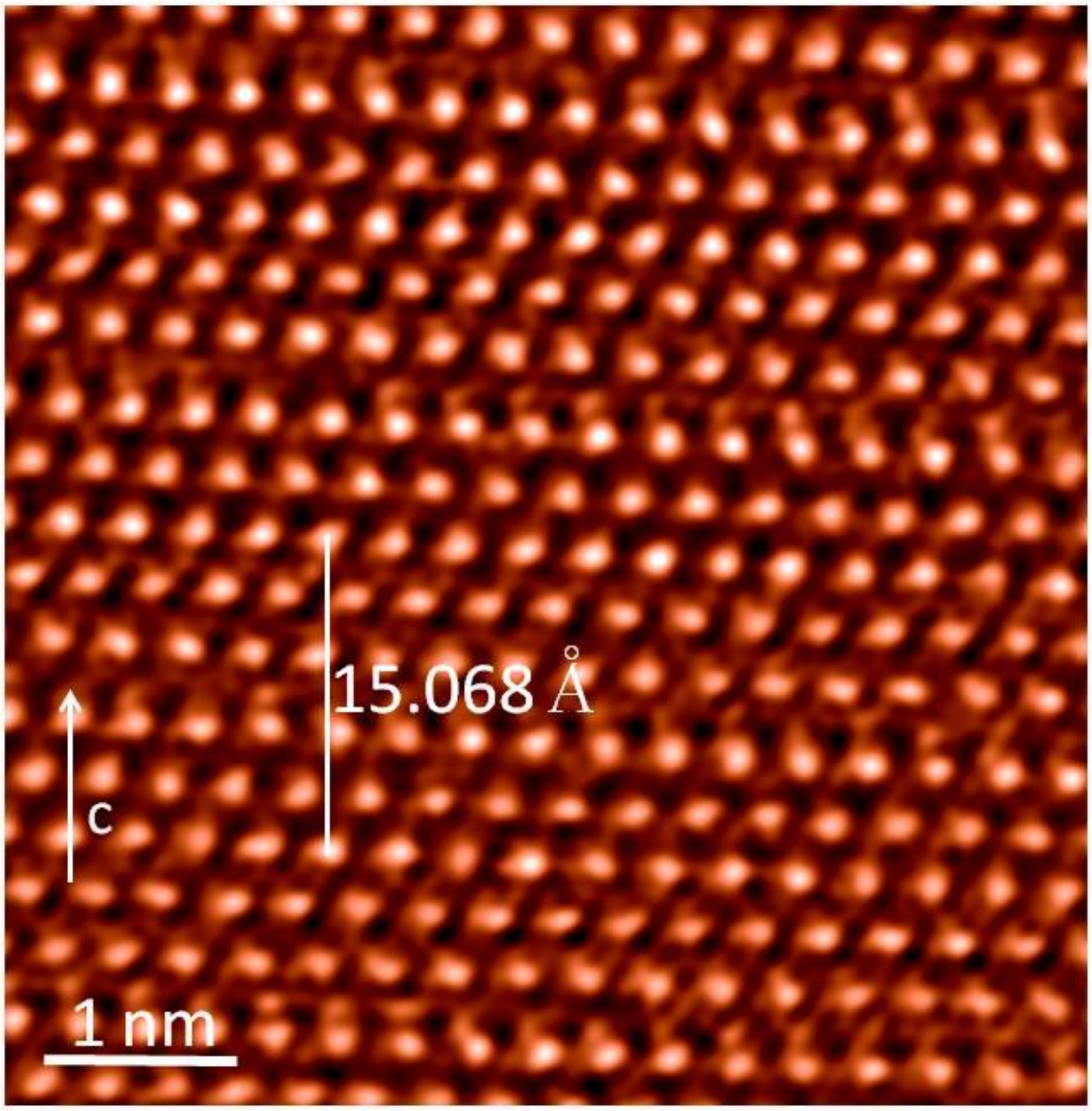

Figure S3. Atomic resolution TEM image of the $Nb_2Pd_xS_{5-\delta}$ fiber taken along the $[1\bar{1}0]$ zone-axis direction. Periodic atomic array along *c*-axis can be found. The lattice constant of *c*-axis is 15.068 Å. The *c*-axis is formed by stacking of double Nb(S) layers sandwiched by Pd(S) layer. The Pd(1) site is fully occupied while the Pd(2) site contains vacancy. The stacking of the Nb layers and Pd layers in $Nb_2Pd_xS_{5-\delta}$ is very similar to that in $Nb_2Pd_{0.71}Se_5$, where the structure is realized with a pseudo-closest packing of the layers in the manner, ABAB...., and insertion of the statistically distributed atoms Pd(2) in the square-planar sites (see Figure 10 in Journal of Solid State Chemistry **57**, 68-81 (1985)).

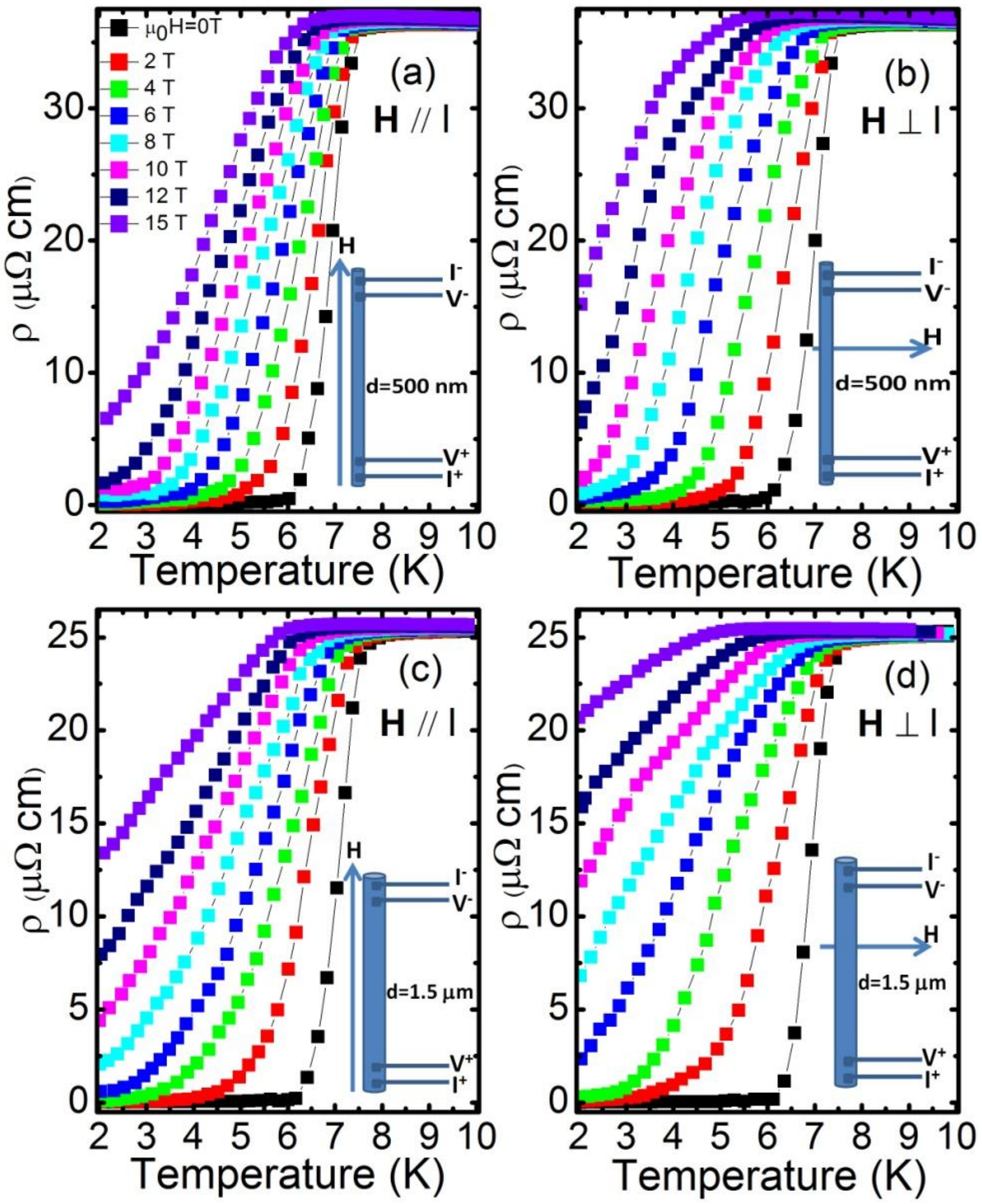

Figure S4. Temperature dependence of resistivity for $Nb_2Pd_{0.963}S_{4.967}$ superconducting fiber under magnetic field of $\mu_0H$=0, 2, 4, 6, 8, 10, 12, and15 T applied parallel and perpendicular to the length of the fibers. The diameters of the fibers are about d=500 nm ((a) and (b)) and d=1.5 μm ((c) and (d)), respectively. The upper critical fields, decided from these ρ~*T* curves, are summarized in the inset of Figure 2. It can be seen that the ratio of the upper critical field with $\mu_0H$//l to that with $\mu_0H\perp$l is only 1.47, suggesting nearly isotropic character of the $Nb_2Pd_xS_{5-\delta}$ superconducting fibers. An interesting fact is that the upper critical field determined from the ρ~*T* curves is strongly dependent on the diameter of the fiber. The upper critical field can be largely enhanced by decreasing the diameter of the fiber. And the upper critical field determined from transport measurement is larger than that determined from magnetization measurement. These results are consistent with previous theory of small-size superconductors (*APL*, **1962**, 1, 7-8; *Phys. Rev.*, **1962**, 127, 386-390; *PRL*, **1962**, 9, 266-267). In a small-size superconductor where the penetration depth is comparable with the size, magnetic field can easily penetrate into the center of a superconductor, thus the diamagnetic signal detected from magnetization measurements should be very weak and the upper critical field determined from magnetization data is not very high. On the contrary, the transport upper critical field could be largely enhanced due to the filamentary structure of the small-size superconductors.

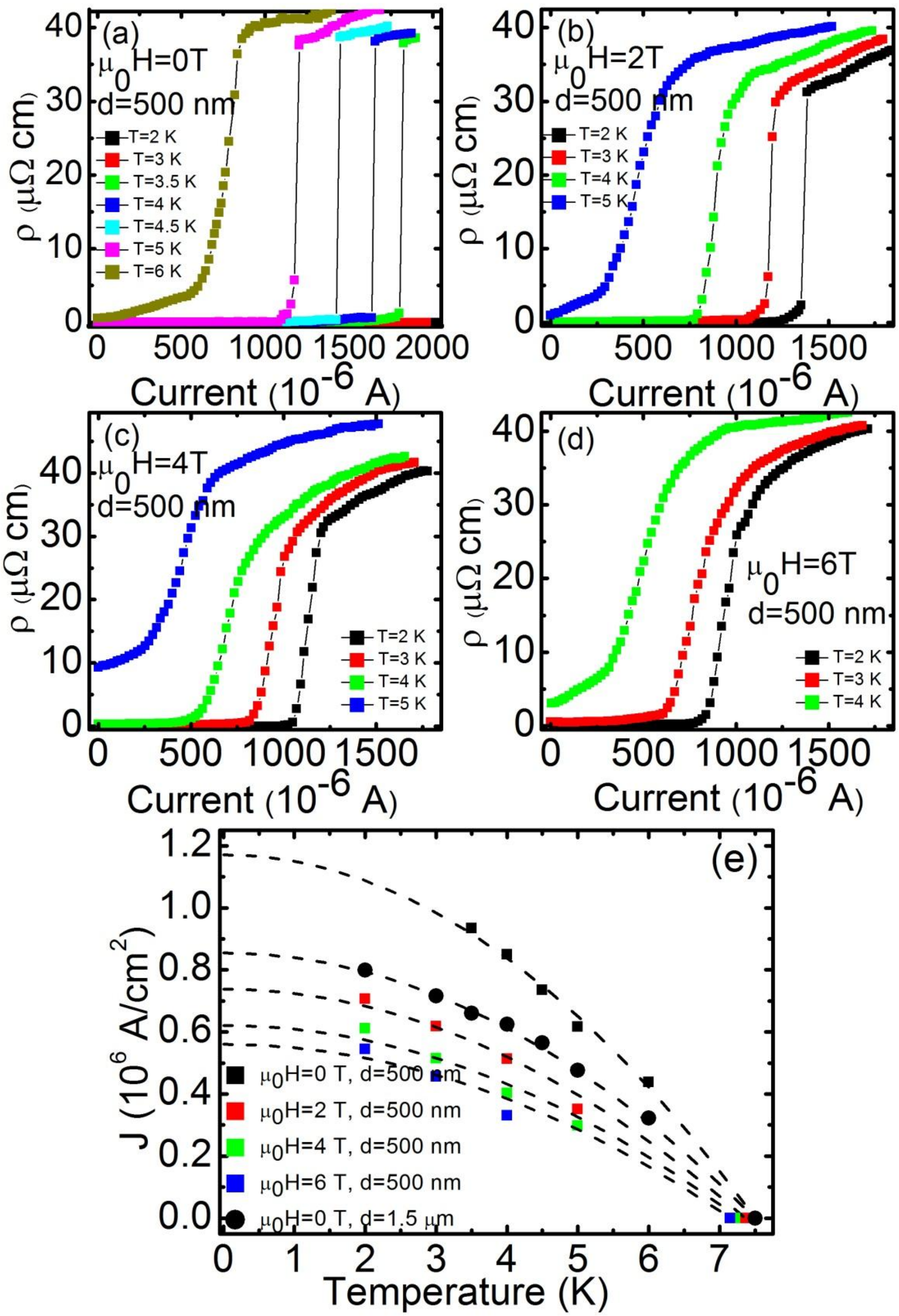

Figure S5. The electric current dependence of resistivity of the $Nb_2Pd_{0.963}S_{4.967}$ superconducting fiber with diameter of d=500 nm measured under (a) 0 T, (b) 2 T, (c) 4 T, and (d) 6 T. The applied magnetic field is along the length direction of the fiber. (e) The critical current density ($J_c$) of the fiber. The dashed curves are the fitting curves using $J_c(T)=J_c(0)\{1-(T/T_c)^2\}$.